\documentclass[
aps,prl
,noshowpacs,
,superscriptaddress
,twocolumn
,twoside
,floatfix
,showemail]{revtex4}
\pdfoutput=1
\usepackage{amssymb,amsmath,graphicx}
\usepackage{bm,mathrsfs,hyperref}
\usepackage[usenames]{color}
\hypersetup{
    colorlinks=true, 
    linkcolor=blue,  
    citecolor=blue,  
}

\begin{document}

\title{Hidden order in hexagonal $\boldsymbol R$MnO$\boldsymbol _3$ multiferroics}
\date{\today}

\author{A. Cano}
\email{cano@icmcb-bordeaux.cnrs.fr}
\affiliation{CNRS, Univ. Bordeaux, ICMCB, UPR 9048, F-33600 Pessac, France }

\begin{abstract}
Hexagonal $R$MnO$_3$ manganites are improper ferroelectrics in which the electric polarization is a by-product of the tripling of the unit cell. In YMnO$_3$, there is a second transition at $\sim$ 920K whose nature remains unexplained. 
We argue that this transition can be seen as a sort of hidden order in which a residual symmetry displayed by the trimerization order parameter is spontaneously broken. This additional order gives rise to twelve structural domains instead of six, and structural domain boundaries that can be either ferroelectric or non-ferroelectric domain walls.
\end{abstract}
\maketitle

Hexagonal $R$MnO$_3$ manganites, with $R=$ Dy-Lu, In, Y, and Sc, were discoverd by Bertaut and collaborators half a century ago \cite{bertaut0}. Nowadays, these compounds are considered as a distinguished class of multiferroic materials \cite{cheong07}. These systems are multiferroics for two reasons. On one hand, ferroelectricity appears at high temperature together with the tripling of the unit cell \cite{bertaut1}. The interplay between the corresponding order parameters gives rise to remarkable features such as clamped ferroelectric-structural domain walls \cite{choi10} with unusual transport properties \cite{meier12}. On the other hand, antiferromagnetic order emerges at low temperatures \cite{bertaut2}, which fits out these systems with additional magnetoelectric properties \cite{me}. 

\begin{figure}[t]
\includegraphics[width=.475\textwidth]{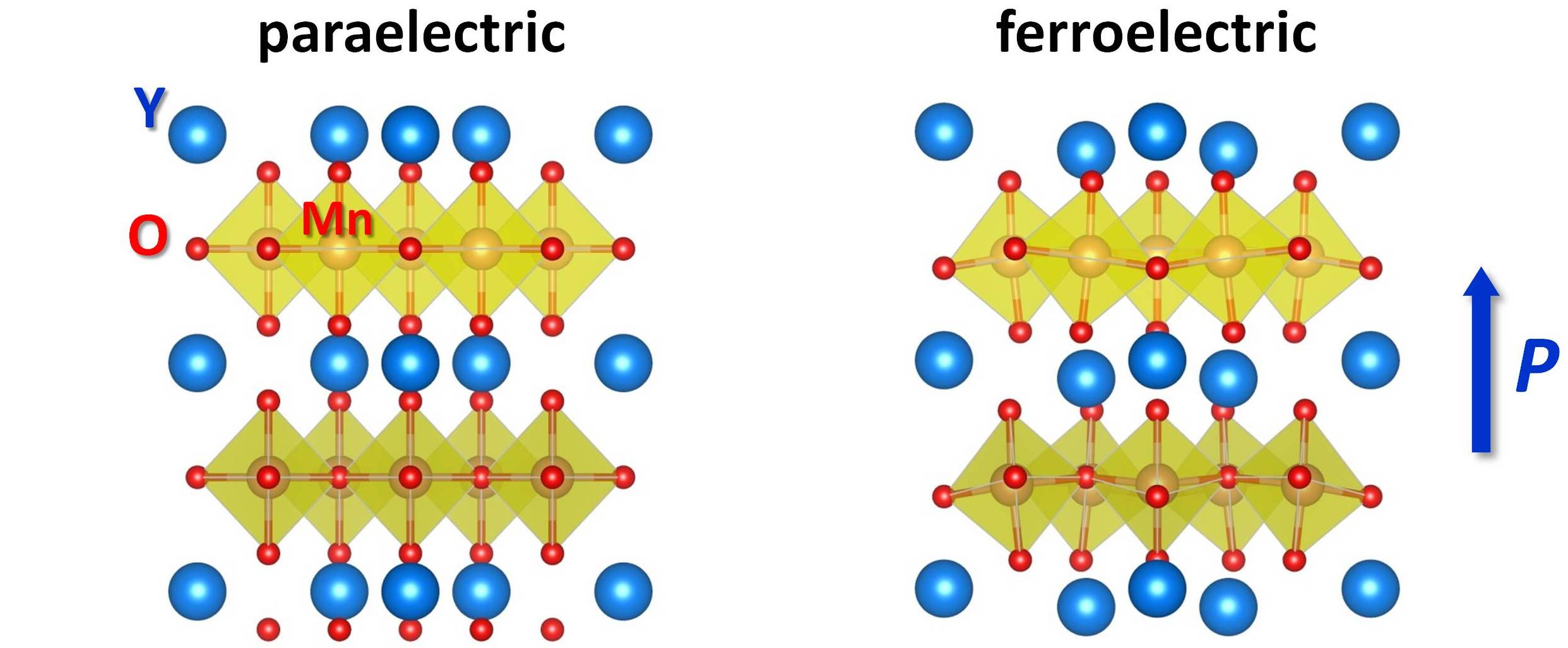}
\caption{Atomic rearrangements associated to improper ferroelectricity in YMnO$_3$, in which the spontaneous polarization $P$ appears along the $c$-axis.}
\label{unit-cell}
\end{figure}

YMnO$_3$ is probably the most studied member of this family. 
There is consensus that YMnO$_3$ is ferroelectric at room temperature, while it has a centrosymetric structure above $\sim$1250K. 
The symmetry of the ferroelectric phase has been ascribed to the $P6_3cm$ space-group. As such, it is connected to the high-temperature $P6_3/mmc$ structure by the tripling of the corresponding unit-cell and the loss mirror symmetry perpendicular to the $c$-axis. This is realized by the tilting and distortion of the MnO$_5$ bipyramids and the displacement of the Y atoms, which triggers the spontaneous electric polarization of the system (see Fig. \ref{unit-cell}). The exact nature of this ferroelectric transition, however, has been the subject of a debate that has intensified during the last decade \cite{intermediate,lonkai04,fennie05}. 

At present, it is widely accepted that trimerization and polarization appear both at once. That is, the symmetry changes in one single step from $P6_3/mmc$ to $P6_3cm$ at 1250K. Thus, the primary order parameter transforms according to the $K_3$ irreducible representation of the $P6_3/mmc$ space group, while the ferroelectricity is a by-product of the structural transition \cite{lonkai04,fennie05}. The system is therefore an improper ferroelectric \cite{levanyuk74,cheong07}.
Intriguingly, a second anomaly has been repeatedly reported around $920$K whose origin remains unexplained \cite{intermediate}. 
The aim of this paper is to provide an explanation that is, in fact, generally valid for any improper ferroelectric.  

\begin{figure*}[t]
\includegraphics[width=.25\textwidth]{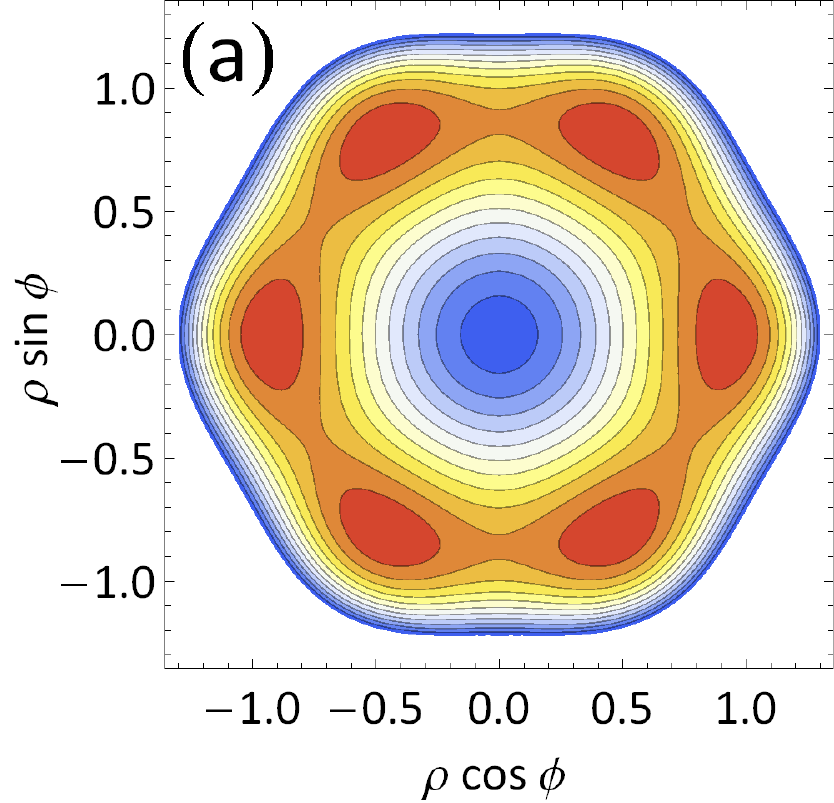}
\hspace{20pt}
\includegraphics[width=.25\textwidth]{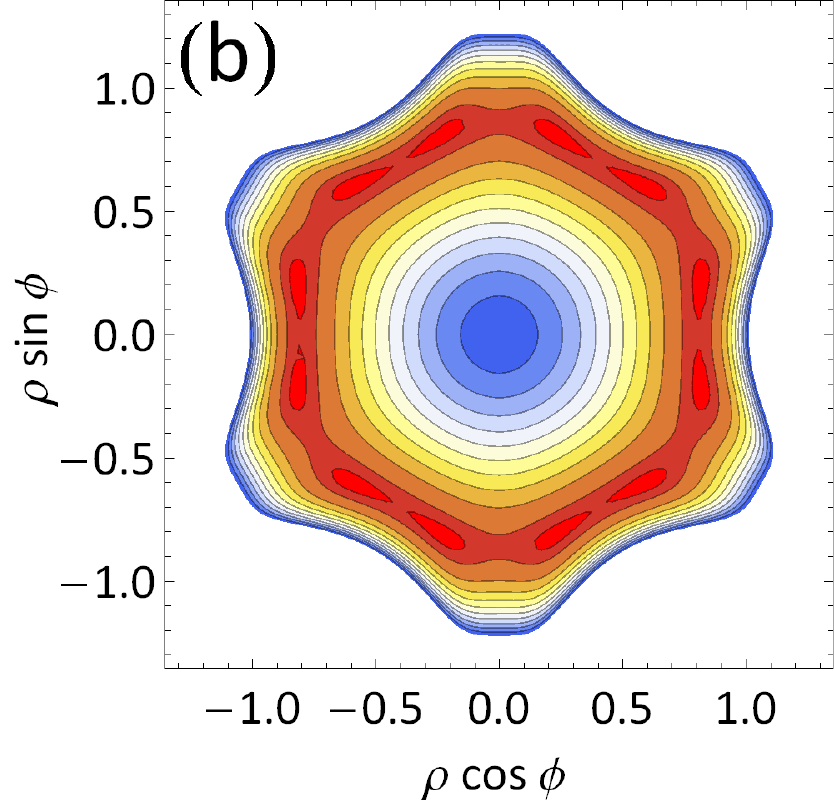}
\hspace{20pt}
\includegraphics[width=.25\textwidth]{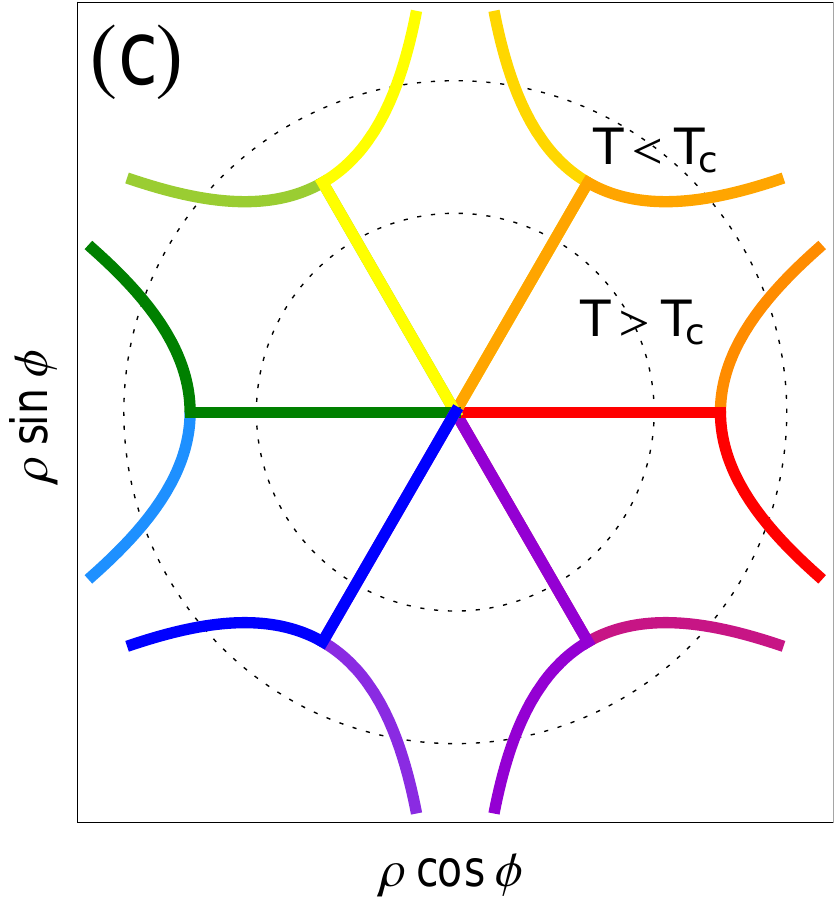}
\caption{(a) Contourplot of the free energy \eqref{F0}. The parameters are taken from \cite{artyukhin12}, which gives rise to six minima (in red). (b) The incorporation of the term \eqref{F12} results into twelve minima ($f=3$ eV \AA$^{-12}$ for the sake of illustration). (c) Illustration of the transition from six to twelve domains as a function of the trimerization amplitude within the model \eqref{F0} and \eqref{F12}. The lines represent the position of the different minima of the free energy. In all cases there are only two polarization domains since $P\sim \cos 3 \phi$.}
\label{energies}
\end{figure*}

Conceptually, this type of secondary transition 
evokes the discussed by Levanyuk and Sannikov in the early days of improper ferroelectrics \cite{levanyuk74,levanyuk70}. Accordingly, it can be seen as a kind of residual symmetry breaking. The point is that the primary $K_3$ order parameter can generate different types of domains in which the symmetry can be broken at different levels. Specifically, the trimerization can be less symmetric than assumed so far (with electric dipoles neither totally compensated nor maximally uncompensated, but somewhere in between). 
The anomalies observed at $\sim 920$K are likely related to the breaking of this residual symmetry. 
We note that the smoking gun for evidencing this phenomenon is not necessarily the electric polarization, as this observable displays only a part of the total symmetry that can actually be broken in the trimerization (i.e. via $K_3$). 

In the case of YMnO$_3$, 
the initial expression of the free energy given by Fennie
and Rabe \cite{fennie05} has been upgraded by Artyukhin {\it et al.} \cite{artyukhin12} taking into account the actual two-component character of the corresponding order parameter $(\phi_1,\phi_2)=(\rho \cos \phi, \rho \sin \phi)$. This upgrade is essential, for example, for describing the topological defects that appear in these systems. The upgraded expression reads
\begin{align}
F=  F_\text{trim} + F_{P} + F_\text{int}.
\label{F0}\end{align}
Here 
\begin{align}
F_\text{trim}& =  {a\over 2} \rho^2 + {b\over 4} \rho^4 + {c \over 6} \rho^6  + {c' \over 6} \rho^6 \cos 6 \phi 
\nonumber\\
&\quad + \text{gradient terms},\\
F_P &={A\over 2}P^2 + \text{gradient terms},
\end{align}
represent independent contributions associated to the trimerization and the electric polarization respectively, while 
\begin{align}
F_\text{int}=  - g P \rho^3 \cos 3 \phi + {g'\over 2} \rho^ 2 P^2 
\label{Fint}\end{align}
describes the interplay between these variables. 
Furthermore, Artyukhin {\it et al.} obtained the parameters of this expression from {\it ab initio} calculations. Interestingly, the energetics of the trimerization is dominated by the interplay \eqref{Fint} as the bare anisotropy $c'$ is positive and much weaker than the eventual anisotropy $\tilde c' = c' - {3 g^2\over 2(A + g' Q^2)}<0 $ obtained after minimization over $P$. The free energy \eqref{F0} is illustrated in Fig \ref{energies}(a). It generates six different low-symmetry states characterized by the discrete values $\phi_n = {n \pi/3}$ ($n=1,2,\dots, 6$) for the phase of the trimerization order parameter. Since $P ~\sim \rho^3 \cos 3 \phi $, there is a finite electric polarization associated to each of these domains that has equal magnitude for even and odd values of $n$ but opposite signs. 

\begin{figure}[b]
\includegraphics[width=.35\textwidth]{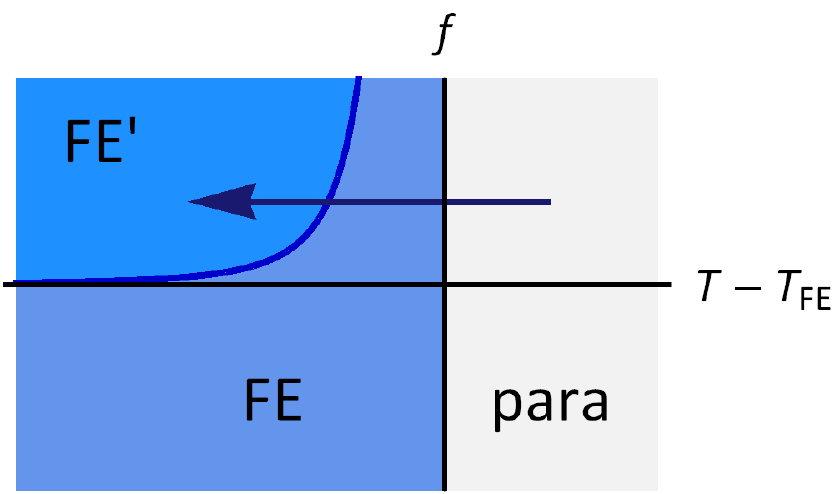}
\caption{Schematic phase diagram for the model \eqref{F0} and \eqref{F12}. As the temperature decreases, the paraelectric phase is replaced by a ferroelectric phase with residual symmetry that can be further broken if $f>0$. The arrow indicates the path likely followed in YMnO$_3$. }
\label{phasediagram}
\end{figure}

In the following we extend this framework in order to reproduce the possibility of a second transition. We note that the above states minimize the free energy by providing the maximal reduction of the anisotropy term ($\cos 6 \phi_n = 1 $). Accordingly, these states can be seen as the most symmetric trimerization states that can be realized in the system. No term in the free energy Eq. \eqref{F0} penalizes this symmetry, which automatically excludes other type of solutions with further reduced symmetry. This situation is contingent, 
and might well be unphysical. The simplest way to correct it is by supplementing \eqref{F0} with the term
\begin{align}
F'={f\over 6} \rho^{12} \cos^4 3 \phi .
\label{F12}\end{align}
If $f>0$, the role of this term is to spoil the preference for symmetric solutions as the trimerization amplitude increases. Thus, the extended free energy eventually describes a new phase in which the six initial domains are split into twelve different states. This is shown in Figs. \ref{energies}(b) and (c). These domains can be characterized by the phases of the trimerization order parameter:
\begin{align}
\phi_{n\pm} = \phi_n \pm \delta, 
\label{extraphi}\end{align}
where $\delta $ is such that 
$\sin 3 (\phi_{n}\pm \delta) = \pm \big(1 - { |\tilde c ' | \over f \rho^6}\big)^{1/2}$. The resulting phase diagram is skectched in Fig. \ref{phasediagram}.

Alternatively, the transition can be seen as driven by the additional order parameter $Q$ that transforms according to the $\Gamma_4^+$ representation of the $P6_3/mmc$ space group. 
At the lowest-order in $Q$, the coupling between this variable and trimerization order parameter is
\begin{align}
F_\text{int}'=\lambda Q \rho^{3} \sin 3 \phi .
\label{Fint'}
\end{align}
Thus, according to \eqref{F0} and \eqref{Fint'}, the emergence of $Q$ also implies the stabilization of additional trimerization states. In this case 
\begin{align}
\sin 3 \phi_{n\pm} = -\lambda {\widetilde c' \over \rho^{6}} Q,
\end{align}
where $Q\sim \pm |T-T_{c}|^{1/2}$ within the Ginzburg-Landau framework. From the point of view of symmetry, the obtained via the high-order term \eqref{F12} is equivalent to this scenario. The physical interpretation, however, is rather different. In the first case, there is no need to invoke any additional order parameter, while the second possibility suggests that there is a sort of hidden $\Gamma_4^+$ 
order behind the second transition in the ferroelectric phase of YMnO$_3$. We note that no $\Gamma$-point phonon corresponds to the $\Gamma_4^+$ symmetry in $R$MnO$_3$ \cite{iliev97}. Consequently, the $Q$ variable itself is associated to orbital degrees of freedom rather than to atomic displacements. 

In both scenarios the additional transition implies a nominal reduction of the magnitude of the polarization since $|P| \sim |\cos 3 \phi_{n\pm}| < 1$. This is tune with the experimental observations. The fact that $P$ eventually reaches a constant value can be understood as the saturation of both the trimerization order parameter and $Q$. 
The residual symmetry breaking describes a $P6_3cm$ to $P3c1$ transition in which both $P\sim \cos 3 \phi$ and $Q\sim \sin 3 \phi$ are non-zero. This is in fact the maximal symmetry breaking that can be obtained via the primary $K_3$ order parameter. It is worth noting that the x-ray diffraction patterns seem to be compatible with such maximally reduced symmetry \cite{bertaut0,rusakov11,kumagai12}. We note that, in general, the temperature evolution can be such that $P$ disappears completely even if the primary $K_3$ order parameter is non-zero. In that case, one recovers six structural domains but with a different $P\bar 3 c1$ space group symmetry. Interestingly, this would have been the case in YMnO$_3$ in the absence of the coupling \eqref{Fint} given that the bare anisotropy coefficient $c'$ is positive, what favors $\sin 3 \phi = \pm 1$ \cite{artyukhin12}. This possibility seems to be realized in InMnO$_3$ \cite{kumagai12}. It would be very interesting to study the effect of the (Y, In) substitution, as the complete sequence of $P6_3cm
\leftrightarrow P3c1 \leftrightarrow
P\bar 3 c1$ transitions between trimerized states could be realized as a function of this doping. 

\begin{figure}[t]
\includegraphics[width=.35\textwidth]{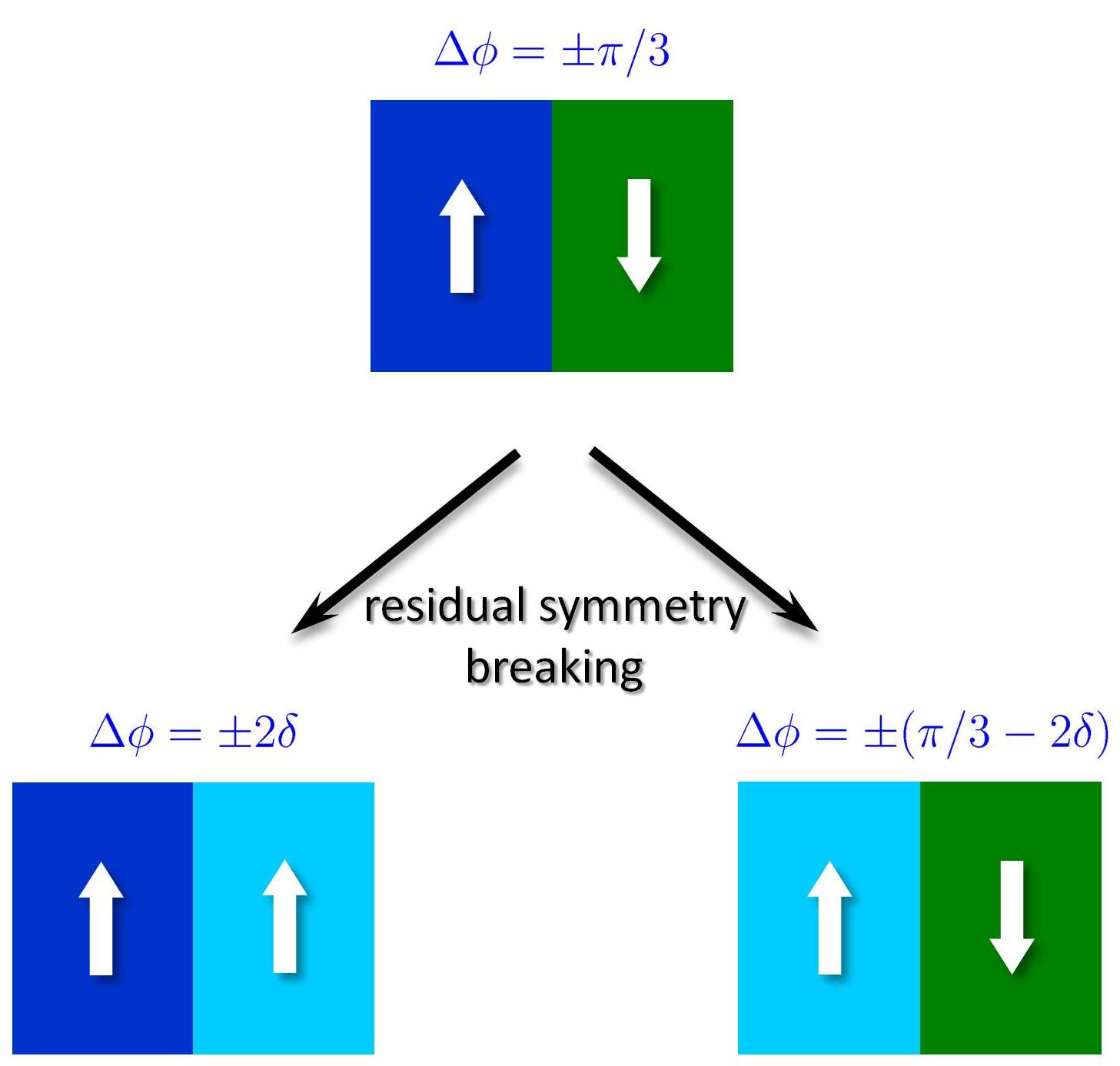}
\caption{Structural domain boundaries expected in YMnO$_3$. (a) If the residual symmetry of the primary $K_3$ order parameter is preserved, the change in the trimerization angle is $\Delta \phi = \pm \pi /3$ and structural and ferroelectric domain walls are interlocked. (b) If the residual symmetry is broken, the change in the trimerization angle can be either $\Delta \phi = \pm (\pi /3- 2 \delta)$ or $\Delta \phi = \pm  2 \delta$, which generates ferroelectric and non-ferroelectric domain walls respectively.  $\delta $ is determined by the amount of residual symmetry that is broken (either by the specific energetics of the trimerization or via a hidden $\Gamma_4^+$ order). }
\label{dw}
\end{figure}

We have seen that the trimerization can give rise to six or twelve different structural domains, depending on whether a residual symmetry is preserved or not. In both cases non-trivial topological defects such as the vortex/anti-vortex pairs observed in \cite{choi10} and the topological stripes discussed in \cite{artyukhin12} can be created. If the residual symmetry is preserved, the topological stability of these defects implies that the change in the trimerization angle across the domain walls is $\Delta \phi = \pm \pi/3$. However, if the residual symmetry is broken, these walls loose their topological stability and each of them will decay into a pair of new walls with $\Delta \phi  = \pm(\pi/3 -2 \delta) $ and  $\Delta \phi  = \pm 2 \delta$. We note that in both cases there are only two types of polarization domains. Thus, while in the first case the structural domain walls are also ferroelectric domain walls, in the second case we have both ferroelectric ($\Delta \phi  = \pm(\pi/3 -2 \delta) $) and non-ferroelectric domain walls ($\Delta \phi  = \pm 2 \delta$) (see Fig. \ref{dw}). This implies that different $C_6$ and $C_3$ vortex/anti-vortex pairs, for example, can look the same if they are probed by means of a technique that reveals the electric polarization only. In all cases, these topological defects will interact differently with electric and strain fields, which can bring additional functionalities to the system. 

\begin{figure*}[b!]
\includegraphics[height=.35\textheight]{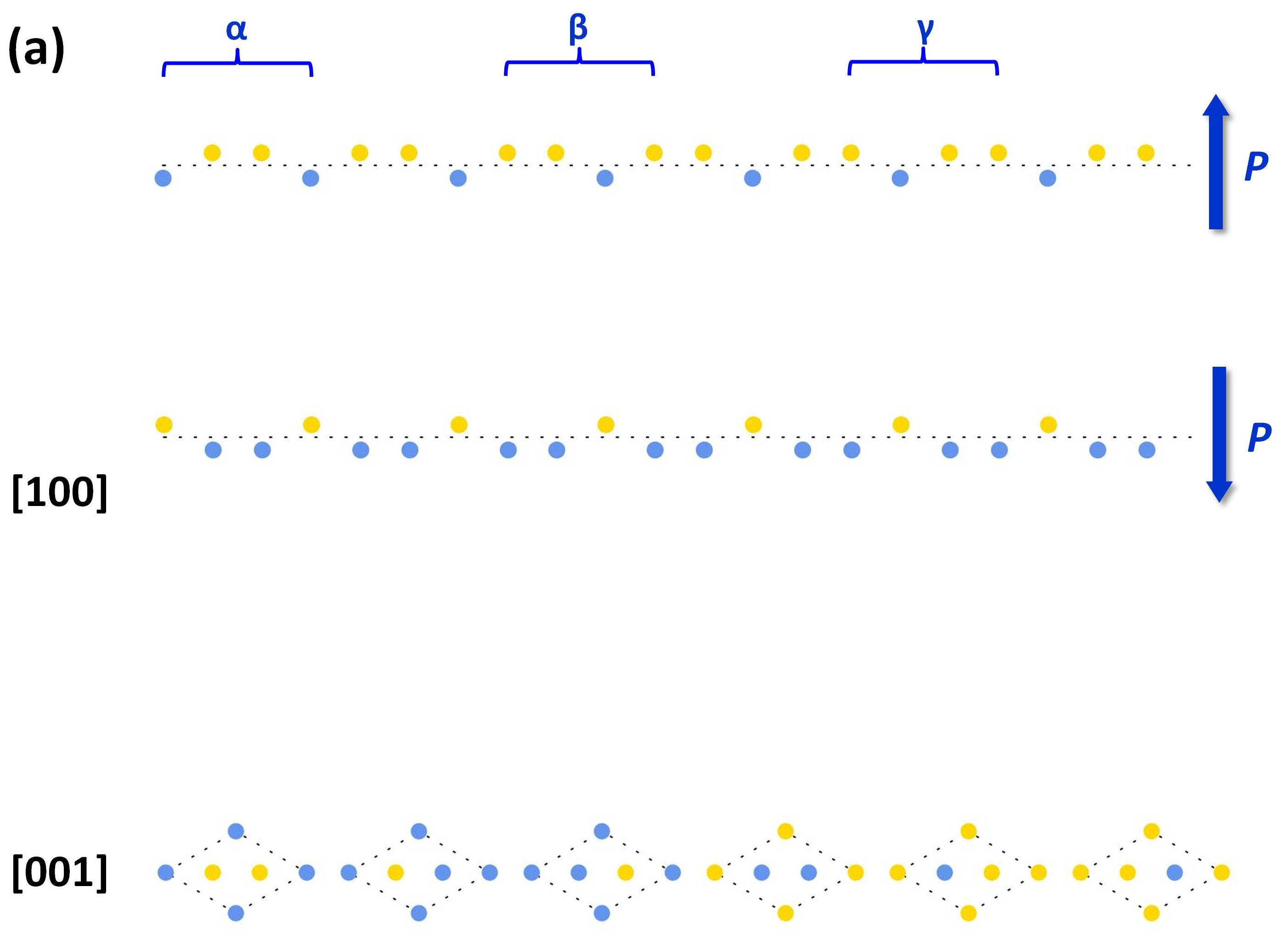}

\vspace{80pt}

\includegraphics[height=.35\textheight]{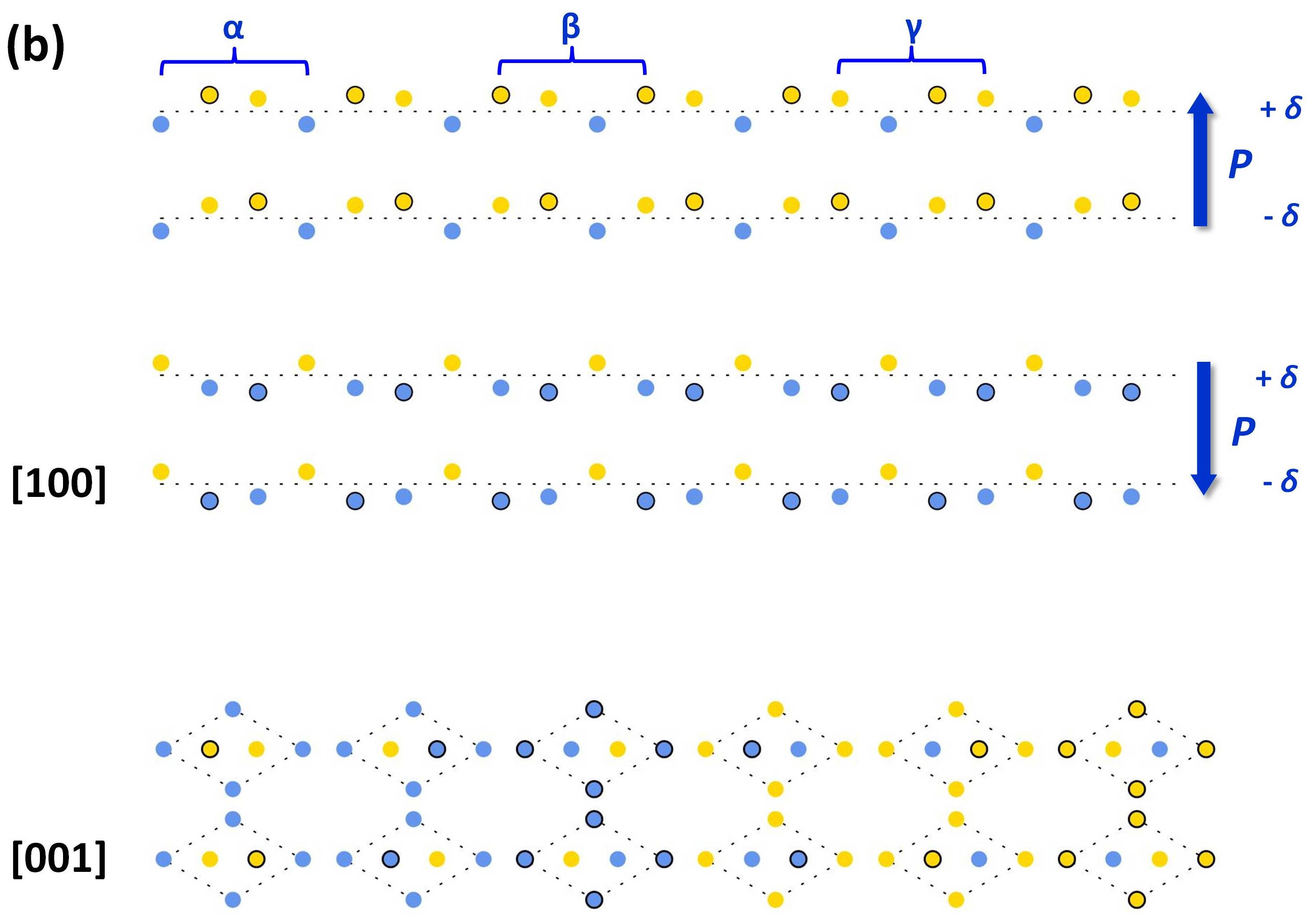}
\caption{Schematic representation of the different ferroelectric $R$ unit cells along the [001] and [100] directions. (a) The residual symmetry of the trimerization order parameter implies two types of $R$ atoms according to their shift, upwards (yellow) or downwards (blue), with respect to their original positions in the paraelectric phase (along the dotted line). The origin of the unit cell determines the structural domains, frequently labelled as $\alpha_ \pm$, $\beta_ \pm$ and $\gamma _\pm$ (where $\pm$ stands for the direction of the electric polarization). (b) The breaking of the residual symmetry generates three inequivalent $R$ atoms, indicated by the additional circled symbols. The number of different unit cells becomes twelve, and the corresponding domains could be labelled $\alpha_{\pm \pm}$, $\beta_{ \pm \pm}$ and $\gamma_{ \pm \pm}$ where the extra $\pm$ stands for the deviation $\pm \delta $ of the trimerization angle from $n\pi/3$ ($n=1,\dots,6$). }
\label{displacements}
\end{figure*}

The atomic patterns associated to the different ferroelectric domains that can appear in $R$MnO$_3$ are illustrated in Fig. \ref{displacements}. If the residual symmetry is preserved, there are two types of $R$ atoms within the [001] layers. 
The electric polarization is due to the relative displacement of these atoms along the $c$ axis, which generates two ferroelectric domains. The structural domains, in their turn, are determined by the different origins of the unit cell. In contrast, if the residual symmetry is broken, the number of inequivalent $R$ atoms becomes three. This increases the number of structural domains, while keeping the two possible directions for the electric polarization. We note that the atomic positions observed in Ref. \cite{zhang13} by means of scanning transimission electron miscroscopy reproduce the second pattern, which confirms that the residual symmetry is broken in YMnO$_3$.

In summary, we have pointed out that the secondary transition systematically observed in YMnO$_3$ is likely related to a residual symmetry breaking of the trimerization order parameter. This can be driven either by the specific energetics of the trimerization or by a hidden $\Gamma_4^+$ order. In both cases, 
twelve different structural domains are obtained out of the six initial states of the system, while the number of ferroelectric domains remains two. Thus, the structural domain boundaries can be either ferroelectric or non-ferroelectric domain walls, which is expected to modify their interaction with the external fields. This enriches the physics of the topological defects characteristic of this type of multiferroics, which can reveal additional functionalities that deserve further studies. 

I thank A. Levanyuk, D. Meier and M. Pollet for useful discussions.

\end{document}